\documentstyle[aps,epsfig]{revtex}
\begin{document}
%
%
%
%
\title{Equilibrium and nonequilibrium fluctuations at the interface between two fluid phases.}
\author{ Pietro Cicuta$^\dag$, Alberto Vailati\footnote{E-mail address:
vailati@fisica.unimi.it} and Marzio Giglio}
\address{Dipartimento di Fisica and Istituto Nazionale per
la Fisica della Materia, Universit\`a di Milano\\ via Celoria 16,
20133 Milano, Italy.\\ $^\dag$ {current address: Department of
Physics, Cavendish Laboratory\\ Madingley Road, Cambridge CB3 0HE,
U.K.}}

\maketitle

%
%
%
\begin{abstract}
We have performed small-angle light-scattering measurements of the
static structure factor of a critical binary mixture undergoing
diffusive partial remixing. An uncommon scattering geometry
integrates the structure factor over the sample thickness,
allowing different regions of the concentration profile to be
probed simultaneously. Our experiment shows the existence of
interface capillary waves throughout the macroscopic evolution to
an equilibrium interface, and allows to derive the time evolution
of surface tension. Interfacial properties are shown to attain
their equilibrium values quickly compared to the system's
macroscopic equilibration time.
\end{abstract}
%
%
%
%
 \vskip 2cm
%


\section{Introduction}
The  fluctuations at the interface between two fluid phases at
thermodynamic equilibrium have been studied very extensively
starting from the beginning of this century  \cite{Man13}, and
particular
 investigation has concerned those at the interfaces of critical
  fluids  \cite{kating68,huwebb69,benedek71,boumeu72}.
  Although the features of equilibrium
  interfacial fluctuations are now relatively well known, the behavior of an
interface under nonequilibrium conditions is still not well
understood.
 Many experiments have been performed to detect an effective
nonequilibrium surface tension in miscible fluids
\cite{joseph,maherprl,petitjeans,maherpre}  but the results
obtained are mainly qualitative.

In this paper we will present experimental results about the
behavior of interfacial fluctuations during the diffusive remixing
of partially miscible phases. In this system the interface between
two fluid phases is being crossed by a macroscopic mass flow. It
is well known that the fluctuations in the equilibrium states
before and after the diffusive partial remixing are controlled by
surface tension and gravity.  It is not clear, however,  what
happens to the interface during the remixing: does the interface
temporarily dissolve? Is there a surface tension during the
transient, and how is it related to the evolving macroscopic
state? We try to address these problems by means of low-angle
light scattering measurements of the correlation function of
fluctuations.\\

The sample considered is a near critical binary mixture kept below
its critical temperature T$_c$, so that it is macroscopically
separated into two bulk phases by a sharp horizontal interface.
The diffusive remixing is started by raising the temperature to a
value closer to T$_c$, but still below it.\\
 We report data at equilibrium showing the expected interface capillary
  waves, and data obtained out of equilibrium during partial
  diffusive remixing, showing for the first time that capillary
   waves at the interface are still present and coexist with
   "giant" nonequilibrium fluctuations in the bulk. By combining
    time-resolved measurements with
predictions for the light scattered by the fluctuations, we are in
the position to derive data for the time evolution of the
nonequilibrium surface tension. These data show for the first time
that during the diffusive remixing the surface tension attains
almost instantaneously its final equilibrium value, and this is
consistent with a  fast rearrangement of the concentration profile
in the neighborhood of the interface.

\section{Experiment}

Traditionally, interfacial fluctuations have been studied by means
of dynamic surface light scattering techniques \cite{langevin}.
These techniques allow to determine the power spectrum of the
light scattered from the excitations, and  have been used very
extensively to characterize the equilibrium properties of
interfacial fluctuations in simple fluids and binary mixtures
\cite{kating68,benedek71,boumeu72,boumeu71}. Surface light
scattering is usually performed by sending a probe beam on the
interface at the angle of total reflection, scattered light being
collected around the specular reflection angle and partially
recombined with the main beam, so that a heterodyne signal is
obtained.  However, dynamic scattering techniques are not very
well suited to study time-dependent nonequilibrium processes, as
the time needed to accumulate an adequate statistics of the
fluctuations is often much longer than the time related to changes
in the macroscopic state of the fluid.  To bypass this problem we
have used a unique low-angle static light scattering setup.
Although the fluctuations' timescales are not accessible with this
instrument, it allows to determine
 in a fraction of a second the static light scattered
at 31 wavevectors distributed within a two decades range, making
it a very useful tool to study fluctuations around a
time-dependent macroscopic state. This instrument, described in
detail elsewhere \cite{cfgpp,vailatigiglio96}, typically
investigates a  wavevector range corresponding to
$100cm^{-1}<q<10000cm^{-1}$.\\
 Our setup is  configured to detect
the transmitted static scattered intensity, the probe beam being
sent vertically at normal incidence. In this way light is
scattered both at the interface and in the bulk layers above and
below it. It can be easily shown that the structure factor in the
transmission geometry is proportional to the usual one in
reflection.\\

A critical binary mixture is an ideal sample to study the
nonequilibrium fluctuations during
 diffusion at an interface. Although other partially
  miscible fluids could be used to
perform this experiment, the use of a critical mixture allows to
tune  the timescale of the macroscopic diffusion process by
adjusting the temperature difference from the critical point.
Moreover experimental runs can be iterated simply by cycling the
temperature.\\ The sample is a 4.5~mm thick horizontal layer of
the binary mixture aniline--cyclohexane prepared at its critical
consolution concentration (c=0.47~w/w aniline). Its critical
temperature T$_c$ is about 30$^o$C, and was determined to within
$\pm$0.01$^o$C before each experimental run. While slowly
 decreasing the temperature  of the single phase above T$_c$,  T$_c$ was
 taken as
the temperature where a  sudden increase  of
 turbidity was observed.\\ The light
scattering cell is a modification of the Rayleigh-B\'{e}nard one
already used to investigate fluctuations in a thermal diffusion
process \cite{vailatigiglio96}, configured to keep the sample at a
uniform temperature. The mixture is sandwiched between two massive
sapphire windows whose temperature control, achieved with Peltier
plates,  is as good as  3mK over a period of one week, temperature
differences between the two plates being kept to about 2mK.\\
 A typical measurement sequence involves the following procedure. An
optical background is recorded with the mixture in its one phase
region at 5.5~K above T$_c$, where the bulk fluctuations amplitude
is many orders of magnitude smaller than the one of equilibrium
and nonequilibrium fluctuations in our experiment. This optical
background mostly contains contributions due to dust and
imperfections of the optical elements, and it is subtracted to all
subsequent measurements. The system is then let phase separate at
3.5K below T$_c$, where the concentration difference between the
two bulk phases is $\Delta c \approx 0.5$\cite{atackrice}. Great
care is dedicated to eliminate wetting drops at the optical
windows. After a few hours the intensity distribution scattered by
the system at thermodynamic equilibrium is recorded, scattered
light being mostly due to the capillary waves at the interface.
The nonequilibrium process is then started by suddenly increasing
the temperature at 0.1$^o$C below T$_c$. The concentration
difference between the phases has to readjust by means of a
diffusive process.  The two macroscopic phases at this temperature
are not completely miscible. At equilibrium they will have
concentration difference of $\Delta c \approx
0.15$\cite{atackrice} and they will be separated by a new
interface. The scattered intensity distribution is recorded during
this transient, until   the system reaches thermodynamic
equilibrium after  about 24~hours. It is well known that the
interfacial fluctuations at the equilibrium states preceding and
following the diffusion process are overdamped capillary waves
\cite{langevin}, characterized by a $q^{-2}$ power spectrum which
exhibits a gravitational stabilization at small wavevectors. The
main problem we want to address is how  the fluctuations behave
during the transient between the equilibrium states. \\

\newpage

\section{Discussion}
 The evolution of the scattered intensity
distributions can be roughly divided into two stages. In the first
stage, represented in Fig.~1, the intensity distribution,
initially due to the equilibrium interface excitations, increases
and develops a bump, which represents the appearance of  a typical
lengthscale. The initial data set, being the least intense, is
particularly affected by the subtraction of the optical
background. The time elapsed during this stage roughly corresponds
to the thermal time needed to increase the temperature of the
sample (about 200~s).
 In the second stage, represented in Fig.~2, the
intensity distribution decreases, until eventually the bump
disappears, and the $q^{-2}$ equilibrium power spectrum of
capillary waves is recovered. Notice that light scattered at large
wavevectors does not change with time during this stage. We will
see that this is related to the readjustment of the concentration
profile across the interface having taken place.\\
 To analyze our data we   assume that the sample may be
depicted as the superposition of two thick bulk layers separated
by a thin interface layer, and that these layers scatter light
independently from each other. Therefore we are considering two
 sources of scattering, the interface fluctuations scattering
 $I_{int}(q)$ and the fluctuations in the bulk phases
scattering $I_{bulk}(q)$, so that the total intensity distribution
is:
\begin{eqnarray}
I(q)=I_{int}(q)+I_{bulk}(q), \label{dis1}
\end{eqnarray}
where, as outlined in the appendix,
\begin{eqnarray}
 \frac{dI_{int}({\bf
q
},t)}{d\Omega\,}\,=\,I_0\,\frac{n^2\,K_0^4}{(2\pi)^2}\,\left(\frac
{
\partial n } {\partial c }\right)^2\,
 \frac{1}{\rho \beta g} k_BT
\,\frac{\Delta c_{int}}{1\,+\,\left(\frac{q}{q_{cap}}\right)^2}
 \label{dis2}
\end{eqnarray}
and
\begin{eqnarray}
 \frac{dI_{bulk}({\bf q },t)}{d\Omega\,}\,=\,I_0\,\frac{n^2\,K_0^4}{(2\pi)^2}\,\left(\frac {
\partial n } {\partial c }\right)^2\, \frac{1}{\rho \beta g} k_BT
\,\frac{\Delta c_{bulk}}{1\,+\,\left(\frac{q}{q_{ro}}\right)^4}.
 \label{dis3}
\end{eqnarray}
In Eqs.~(\ref{dis2}) and (\ref{dis3}), $K_0$ is the wavevector of
light in vacuum, $n$, $c$, $\rho$ are the mixture's refraction
index, weight-fraction concentration and density, respectively.
$\beta$ is $\frac{1}{\rho}\frac{\partial \rho}{\partial c}$ and
$g$ is the gravity acceleration.\\ $\Delta c_{int}$ is the
concentration difference across the interface and $\Delta
c_{bulk}$ is the total sample concentration difference minus
$\Delta c_{int}$, that is the concentration difference that falls
in the bulk phases. The rolloff wavevectors $q_{cap}$ and
$q_{ro}$, given by
\begin{eqnarray}
 q_{cap}\,=\,\left[\frac{\Delta\rho\,g}{\sigma}\right]^{\frac{1}{2}}
 \label{qcap}
\end{eqnarray}
and
\begin{eqnarray}
q_{ro}\,=\,\left[\frac{\beta\,\partial_z
c\,g}{2\,\nu\,D}\right]^{\frac{1}{4}},
 \label{qro}
\end{eqnarray}
characterize the onset of  gravitational stabilization  at large
lengthscales  of capillary and bulk fluctuations, respectively
\cite{Man13,vailatigiglio96}. In Eq. (\ref{qro}) $\partial_z c$
represents the largest concentration gradient in the bulk phases
at a certain time \cite{vailatigiglio97}.

We  use Eqs.~(\ref{dis1})-(\ref{dis3}) to fit the experimental
data shown in Figs.~1--2. In order to limit the number of fitting
parameters we take advantage of the fact that the concentration
difference $\Delta c$ across the whole sample does not change
during the so called free-diffusive regime \cite{vailatigiglio97,vailatigiglio98},
as the diffusive remixing initially involves
only layers of fluid close to the interface. Therefore $\Delta
c_{int}$ and $\Delta c_{bulk}$ are related by
\begin{eqnarray}
\Delta c_{int}(t=0)\,=\,\Delta c_{int}(t)\,+\,\Delta c
_{bulk}(t)\,\,\,\,\,\,\,\,\,\,\,\,\,\,\,\,\, t\,\ll\,\tau_{macro}
 \label{dis4}
\end{eqnarray}
where $\tau_{macro}=\frac{s^2}{\pi^2\,D}$ is  the time required
for diffusion to occur over the sample height $s$. $\tau_{macro}$
is of the order of~7000s for our sample, by assuming D=$6\,\cdot
\,10^{-7}\frac{cm^2}{s}$ (this is the
 equilibrium value at 3\,K below T$_c$, see \cite{vailatigiglio97}
 and references therein). By imposing the reference value $\Delta
c_{int}(t=0)=0.5$ \cite{atackrice} and by fitting the experimental
data using Eqs.~(\ref{dis1})-(\ref{dis3}), we are able to
determine three parameters: the concentration difference across
 the interface and the rolloff wavevectors $q_{cap}$ and $q_{ro}$.\\
\begin{figure}[f]
          \begin{center}
          \fbox{\epsfig{file=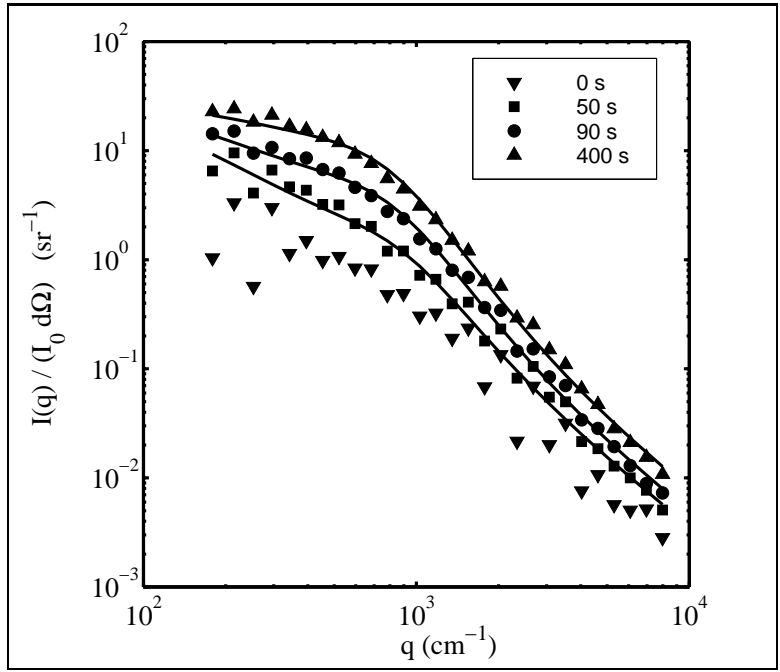,width=10cm}}
          \end{center}
        \caption{Normalized scattered intensity plotted vs. scattered
wavevector $q$ at  different times. The initial
dataset~($\bigtriangledown$) is the intensity scattered by the
initial equilibrium interface. Time is measured from the start of
the partial remixing process. The figure shows the early stages of
the remixing process, when the scattered intensity is increasing
with time. The solid lines represent the best fit of the
experimental data with Eqs.~(\ref{dis1}-\ref{dis3}).} \label{F1}
\end{figure}
\begin{figure}[f]
          \begin{center}
          \fbox{\epsfig{file=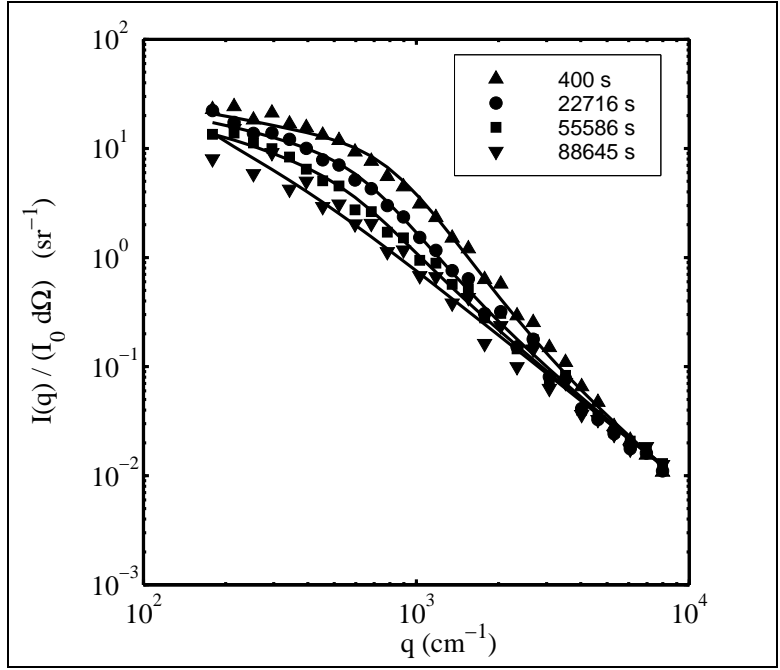,width=10cm}}
          \end{center}
        \caption{Normalized scattered intensity plotted vs. scattered
wavevector $q$ after the onset the remixing process. The scattered
intensity is decreasing with time. The least intense
scattering~($\bigtriangledown$) is due to the final equilibrium
interface. The solid lines represent the best fit of the
experimental data with Eqs.~(\ref{dis1}-\ref{dis3}). } \label{F2}
\end{figure}
\clearpage
  Results for
$\Delta c_{int}$ are presented in Fig.~3(a) as a function of time.
\begin{figure}[h]
          \begin{center}
          \fbox{\epsfig{file=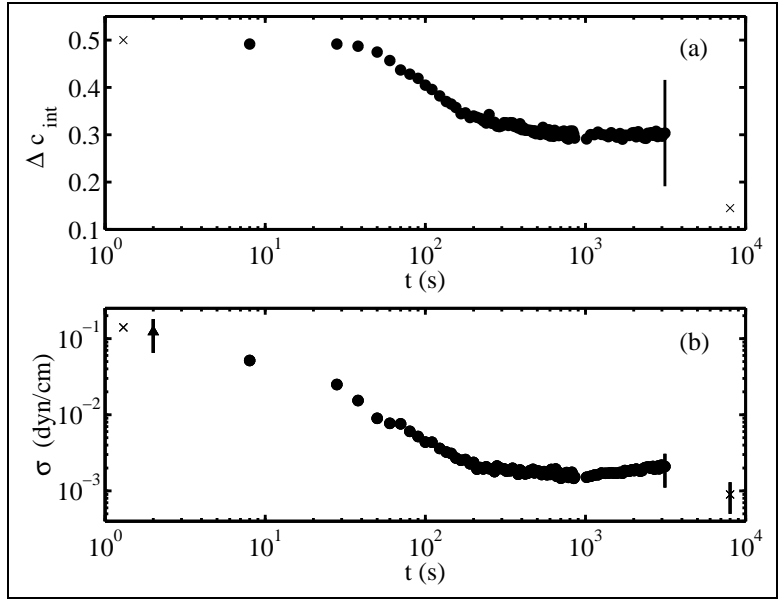,width=10cm}}
          \end{center}
       \caption{ Time evolution of the concentration difference across
the interface, (a), and of the interfacial surface tension, (b).
Circles represent the experimental results obtained by fitting the
scattered intensity distributions with
Eqs.~(\ref{dis1}-\ref{dis3}).  Crosses mark the reference
equilibrium values in the initial and final states. The triangle
in panel (b) corresponds to the surface tension measured in the
initial equilibrium state. Vertical bars on the experimental
results represent the estimated fitting and sistematic errors.}
\label{F3}
\end{figure}
Initially $\Delta c_{int}$ has its equilibrium value $\Delta
c_{int}=0.5$. After about 40~seconds from the temperature increase
it begins to drop, and decreases for about 300s, finally
stabilizing to the constant value $\Delta c_{int}=0.3$. This is
roughly a factor of two larger than the reference equilibrium
value \cite{atackrice}. This discrepancy is due to the difficulty
(both theoretical and in the fitting procedure) of clearly
ascribing the measured scattered intensity to either interface or
bulk phases, and an
 estimate of our rather large fitting and systematic
 error is shown with the error bars in Fig.~3. This error is
  such that our data cannot be considered fully quantitative, but it
   does not affect the features we discuss.\\
 By combining the
results for $\Delta c_{int}$ and $q_{cap}$, since from Eq.
~(\ref{qcap})
\begin{eqnarray}
\sigma\,=\,\frac{ \rho \beta \Delta c_{int} g }{q_{cap}^2},
 \label{dis5}
\end{eqnarray}
 we are in the
unique position to obtain the time evolution of the interfacial
surface tension during the nonequilibrium process.
 Experimental results for the surface tension are
shown in Fig.~3(b), which represents the main accomplishment of
this work. The two crosses
mark the value of the equilibrium surface tension at the initial
and final temperature, extrapolated from the reference data from
Atack and Rice \cite{atackrice}. The first data point represents
the equilibrium surface tension measured with our light scattering
setup. The agreement with the reference value is  good. After the
diffusion process is started the surface tension drops about two
orders of magnitude, until after about 300s it stabilizes to a
constant value. The asymptotic value of the surface tension is
about a factor of two larger than the reference value, a good
result considering the wide range of values spanned. Although
these results are only partially quantitative, Fig.~3
unambiguously shows for the first time that the properties of the
nonequilibrium interface rapidly attain their equilibrium values.
This equilibration time is very small compared with the one
associated to readjustments of the bulk phases (which corresponds
to about one day), and it is comparable to the time needed to
increase the temperature of the sample. Notice that the surface
tension evolution does not show the initial delay seen on the
$\Delta c_{int}$
 evolution. This indicates that the surface tension is probably
 following the local temperature almost instantly, whereas
 $\Delta c_{int}$ does not change until diffusion has occurred
  over the fluctuations' characteristic lenghtscales. With the
  diffusion coefficient given above this time is about 30s for
   the smallest wavevectors observed.\\

We are now in the position to comment the fast growth, shown in
Fig.~1, of the scattered intensity distributions at intermediate
and large wavevectors. Soon  after the diffusion process is
started the concentration difference across the interface
decreases at its equilibrium value. According to the concentration
conservation Eq.~(\ref{dis4}), a strong concentration difference
$\Delta c _{bulk}$ is rapidly created in the bulk phases, and a
large concentration gradient quickly grows
 near the interface.  This gives rise to
velocity-induced concentration fluctuations described by
Eq.~(\ref{dis3}) \cite{vailatigiglio97,vailatigiglio98}.  The
growth of $\Delta c_{bulk}(t)$ during the free diffusive regime is
shown by the circles in Fig.~4, and it mirrors the results for
$\Delta c _{int}$ in Fig.~3(a). When enough time has passed for
diffusing particles to reach the macroscopic boundaries (about
5000s as from Fig.~4), the sample enters the restricted diffusion
regime, where the total concentration difference across the sample
begins to change, and Eq.~(\ref{dis4}) does not hold any more.
However, according to Fig.~3, we can now assume that the
interfacial parameters have attained their asymptotic values. In
this way we can fit the scattered intensity distributions to
determine the concentration difference across the bulk phases.
Results for $\Delta c _{bulk}$ obtained in this way are shown by
the squares in Fig.~4.
\begin{figure}[h]
          \begin{center}
          \fbox{\epsfig{file=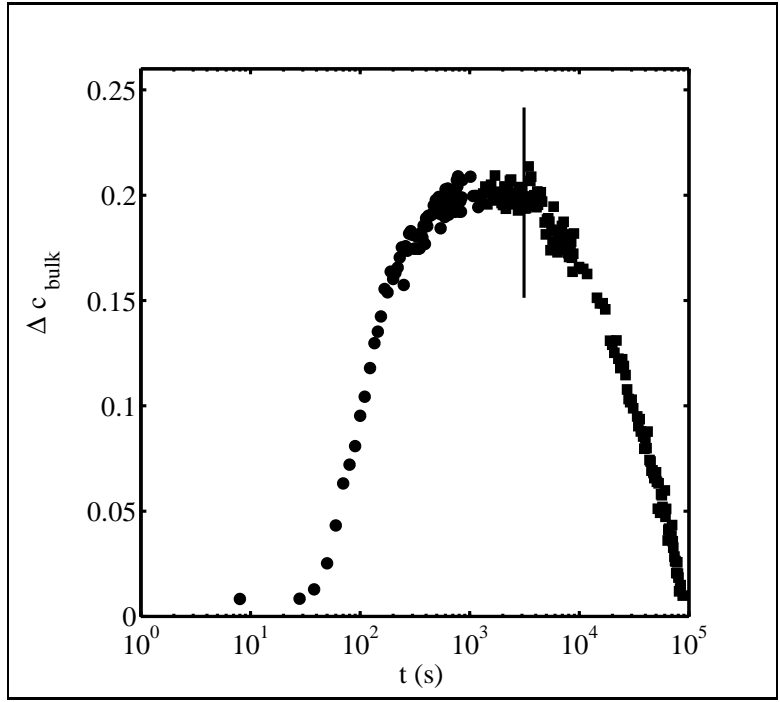,width=10cm}}
          \end{center}
       \caption{Time evolution of the concentration difference across the
bulk phases. During the initial free-diffusion stage (circles) the
total concentration across the sample is conserved (see
Eq.~(\ref{dis4})) and the data mirror those presented in Fig. 3(a)
for the concentration difference across the interface. Later on
the presence of boundaries is felt, and the sample enters the
restricted-diffusion stage (squares). In agreement with  the
results presented in Fig.~3,  we have assumed that during this
stage the interfacial properties have already attained their
equilibrium values. } \label{F4}
\end{figure}
The evolution of this process is very similar to that already
observed during the free diffusion of completely miscible phases
\cite{vailatigiglio97,doriano}. From the fitting of the
nonequilibrium bulk data we  also  determine the rolloff
wavevector $q_{ro}$, which corresponds to the bump observed in the
light scattering data in Figures~1-2. As outlined above and
thoroughly described in Ref.\cite{vailatigiglio98}, fluctuations
at wavevectors smaller than $q_{ro}$ are stabilized by gravity,
which frustrates the $q^{-4}$ divergence at small wavevectors. The
rolloff wavevector is plotted in Fig.~5 as a function of time.
\begin{figure}[h]
          \begin{center}
          \fbox{\epsfig{file=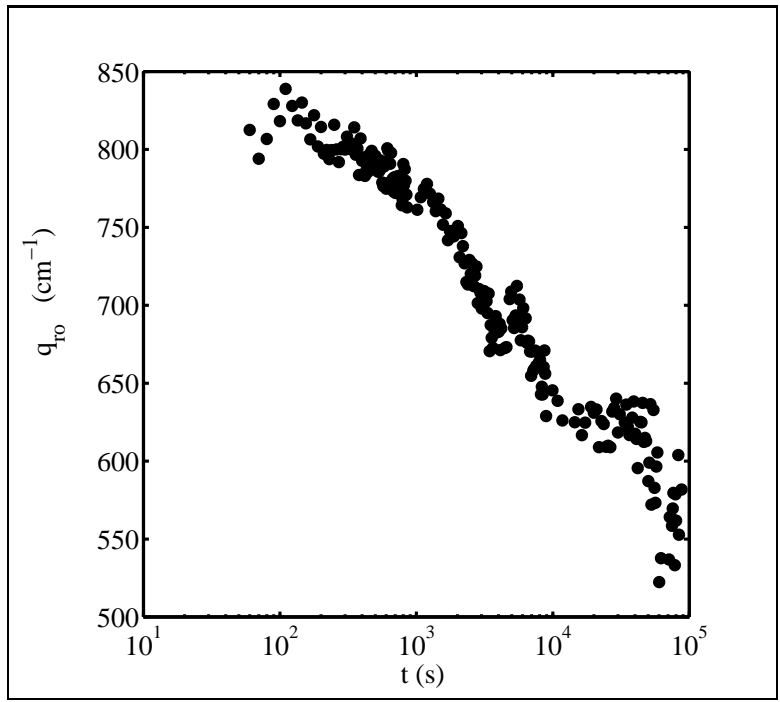,width=10cm}}
          \end{center}
       \caption{Time evolution of the rolloff wavevector $q_{ro}$. }
\label{F5}
\end{figure}
The results displayed in Fig.~5 show that the variation of
$q_{ro}$ roughly corresponds to a factor of~1.4. $q_{ro}$ is
mostly determined by the layers of fluid where the concentration
gradient is largest \cite{vailatigiglio97}, that is the bulk
layers close to the interface. From Eq.~(\ref{qro}), we can
estimate that the variation of the concentration gradient close to
the interface is rather small, roughly corresponding  to a factor
four. This behavior suggests that the bulk concentration gradient
is pinned to the interface concentration profile.

\section{Conclusions}
We have reported light-scattering measurements of the correlation
function of fluctuations at the interface and in the bulk phases
 of a critical binary mixture undergoing diffusive partial remixing.
 From the intensity distributions it is clear that even during remixing
  a sharp interface separates the two bulk phases,
 and that this interface is roughened by capillary waves similar
  to those at equilibrium. Our data have been analyzed,
  yielding qualitative information on the concentration profile
   of the system. In particular our main results are the time
   evolution of the interface concentration difference and the
 surface tension, showing how these parameters
   relax to the final equilibrium values quickly compared to
   the system's macroscopic equilibration time.\\

 \acknowledgments
   We thank Dr. Giuseppe Gonnella for useful comments. Work partially supported by the
   Italian Space Agency (ASI).

\appendix
\section*{}
   In this appendix we shall  outline how one may derive the  structure factor
   of fluctuations in an out of equilibrium binary mixture where a
concentration gradient and a diffusive concentration flux are
present. We consider a fluid described by a $z$ dependent
concentration gradient $c(z)$. Layer by layer, surfaces of uniform
concentration can be defined. From a macroscopic point of view
these surfaces are horizontal planes. In the presence of
fluctuations the surfaces are corrugated, due to the motion of
parcels of fluid in the vertical direction. As customary, we will
indicate by $h_z(x,y,t)$ the vertical displacement of the surfaces
from their mean position at $z$.\\
In such a system   light scattered with scattering vector {\bf q}
is proportional to the mean square amplitude of the roughness mode
 {\bf q}, $\langle \left| h_{{\bf q}}(t) \right|^2
\rangle$. We shall show how this quantity may be calculated for
the two limiting cases of a sharp interface and a linear
concentration gradient, extending the fluctuating hydrodynamics
treatment  applied in \cite{vailatigiglio98}.\\
 Throughout this
paper we are dealing with small fluid velocities and overdamped
motion, and equations will be approximated accordingly, see for
example the discussion in \cite{langevin}. We will consider the
scattering of light by long-wavelength fluctuations, which scatter
light mostly at small scattering angles in the forward
direction.\\ We shall suppose fluctuations to be generated
independently at different heights, and shall first consider a
single fluctuation generated in a layer $d z$ at $\bar{z}$. Then,
for every wavevector {\bf q}, one has that $h_z({\bf q})$ will be
maximum for $z=\bar{z}$ where the fluctuation was generated, and
exponentially decreasing with distance from $\bar{z}$ like
$h_z=h_{\bar{z}}\exp(-q|z-\bar{z}|)$. This is a consequence of the
hydrodynamic equations for a viscous fluid in the case of
overdamped motion \cite{langevin}.\\
    The hydrodynamic equations describing
motion in an incompressible viscous mixture can be linearized for
small velocities \cite{vailatigiglio98} as:
\begin{eqnarray}
 div\,{\bf v}\,=\,0
 \label{divvzero}
\end{eqnarray}
\begin{eqnarray}
 \frac{\partial\,{\bf v}(x,y,z,t)}{\partial\,t}\,=\,
\nu\,\nabla^2\,{\bf v}\,-\,\frac{1}{\rho}\,\nabla\,P(x,y,z,t)
 \label{motointerfaccia}
\end{eqnarray}
\begin{eqnarray}
 \frac{\partial\,c}{\partial\,t}\,=\,-{\bf v}\cdot\langle\nabla
 c\rangle_t\,-\frac{1}{\rho}\,\nabla\cdot{\bf j}
 \label{convdiff}
\end{eqnarray}
where ${\bf v}$ is the fluctuating velocity, $c$ the fluctuating
concentration, $\nu$ is the kinematic viscosity, ${P}(x,y,z,t)$ is
the pressure, $\langle \nabla c \rangle_t$ is the
concentration gradient averaged over typical fluctuations'
timescales and ${\bf j}$ is the mass diffusion flux.
 Equation~(\ref{divvzero}) is the usual mass conservation
equation for an incompressible fluid,  Eq.~(\ref{motointerfaccia})
is the linearized Navier--Stokes equation, and
Eq.~(\ref{convdiff}) is the convection-diffusion equation. The
macroscopic diffusive flux ${\bf j}$ is proportional to
$\nabla\mu$, where $\mu$ is the chemical potential of the mixture;
it is non--zero where $\mu$ is not  constant.\\ Equiconcentration
surfaces are corrugated by
 thermal velocity fluctuations in the vertical
direction, and these are described by the $z$~component of
Eq.~(\ref{motointerfaccia}).\\ We change variable from $c$ to $h$
in Eq.~(\ref{convdiff}), and simplify the
 equation  by assuming that concentration diffusion obeys Fick's law
$\frac{\partial h}{\partial t}=D\nabla^2 h$, where $D$ is the
"mutual diffusion" coefficient.\\ From Eq.~(\ref{convdiff}) we
take the $z$ component of the velocity,
$v_z\,=\,\frac{\partial}{\partial\,t}h\,+\,D\nabla^2 h$, and
substitute it into equation~(\ref{motointerfaccia}), obtaining,
after Fourier transform in $x,y$ and $t$,  an equation of motion
for non-equilibrium fluctuations:
\begin{eqnarray}
 -i\,\omega\,\left[\,-i\,\omega\,h_{{\bf q},\omega}(z)\, + \,2D\,q^2\,h_{{\bf q},\omega}(z)
 \,\right]\,&=&\nonumber\\=\,
2\,\nu\,q^2\,\left[\,-i\,\omega\,h_{{\bf
q},\omega}(z)\,+\,2D\,q^2\,h_{{\bf q},\omega}(z)
\,\right]\,-\,\frac{1}{\rho}\,q\,P_{{\bf
q},\omega}(z)\,&+&\,\frac{1}{\rho}\,q\,S_{{\bf
q},\omega},\nonumber\\
 \label{motointerfacciah2diff}
\end{eqnarray}
where  the continuity of  tangential stresses on the fluctuating
surface has been imposed and a stocastic force term $S_{\bf q}$
has been added to describe the onset of thermal spontaneous
velocity fluctuations. The correlation function of this stocastic
force is assumed to be (see~\cite{fischliq})
$\langle\left|\frac{1}{\rho}{q}{ S}_{{\bf
q}\,\omega}\right|^2\rangle\,=\,k_BT\frac{2\,\nu\,q^3}{\rho\,A}$
as in equilibrium.\\
 The pressure {\em P} in Eq.~(\ref{motointerfacciah2diff}) may be exerted by the external
gravity force and by internal capillary forces and is measured
against the average fluid pressure. Because of Eqs.
(\ref{divvzero})-(\ref{motointerfaccia}), in deriving
  Eq.~(\ref{motointerfacciah2diff}) we have imposed that
  the pressure $P(\bar{z})$ induced by the fluctuation in layer~$\bar{z}$
  decays exponentially with
  distance from the fluctuation layer  to the average pressure,
  like $P_{\bf q}(z) \,=\,P_{\bf
q}(\bar{z})\,\exp\left({-\,q\,|z-\bar{z}|}\right)$. This pressure
term depends on the local concentration profile. It can be written
explicitly  for fluctuations involving a "sharp interface" profile
and for those in the  bulk phases, by considering the gravity and
capillary forces  on the fluctuations.
 We shall consider a fluctuation to be a sharp-interface fluctuation if
  $\bar{z}$ is close to the interface position $z_i$. Instead, if $\partial_z
 c(z)$ may be considered linear  within a range of $\pm\frac{1}{q}$
 around $\bar{z}$, the fluctuation will
 be considered a bulk fluctuation. Our
treatment is approximate in that we are supposing that every
fluctuation falls within one of these cases.\\ Interface
fluctuations and bulk fluctuations give rise to different
dynamics.\\ In the case of a sharp interface, the pressure acting
on a fluid element at height $\bar{z}=z_i$, when a fluctuation
occurs at $z_i$ bending the interface, is:
\begin{eqnarray}
P_{\bf q}(z_i) \,=\,\frac{1}{2}\,\left[\,\Delta\rho\,g\,h_{\bf q
}(z_i)\,+\,\sigma\,q^2\,h_{\bf q }(z_i)\right].
 \label{frozagravcapb}
\end{eqnarray}
Substituting this pressure in Eq. (\ref{motointerfacciah2diff})
one may, with algebra similar to that in \cite{vailatigiglio98},
calculate the spectrum of interface fluctuations:\\
\begin{eqnarray}
 \left|h_{{\bf q}\,\omega}\right|^2\,=
 \,{\langle\left| h ({\bf q})\right|^2\rangle}_t\,
\frac{2 \frac{
 \Delta\rho\,g\,q\,+\,\sigma\,q^3\,+\,4\,\rho\,\nu\,D\,q^4
}{4\,\rho\,\nu\,q^2}} { \omega^2\,+\,\left(\frac{
\Delta\rho\,g\,q\,+\,\sigma\,q^3\,+\,4\,\rho\,\nu\,D\,q^4
}{4\,\rho\,\nu\,q^2}\right)^2},
 \label{motointerfaccia5diff}
\end{eqnarray}
 where the static term is:
\begin{eqnarray}
 {\langle\left| h ({\bf q})\right|^2\rangle }_t\,= \,\frac{1}{A}\frac{k_BT}{\Delta\rho\,g\,+\,\sigma\,q^2\,+\,4\,\rho\,\nu\,D\,q^3}.
 \label{motointerfaccia6diff}
\end{eqnarray}
At equilibrium, when the mutual diffusion coefficient $D\,=\,0$,
the spectrum linewidth is the same as that of the well-known
equilibrium interface \cite{boumeu72,langevin}: gravity dominates
at small wavevectors and capillary forces at large ones. However,
during the nonequilibrium process, diffusion is effective in
relaxing large--wavevector fluctuations. At these wavevectors the
linewidth becomes the usual diffusive $D q^2$ one, and this
diffusive contribution is also apparent
  in the static structure factor where a
new term proportional to $q^3$ is present in the denominator. With
our setup it was not possible to study the structure factor at
large enough wavevectors to check the result and, as far as we
know, this feature  has never been observed.\\
 The other case is
that of a bulk phase where the concentration gradient is
approximately linear. Here the pressure opposing a fluid element
at height $\bar{z}$ when a fluctuation  occurs bending the layer
at height $\bar{z}$ is:
\begin{eqnarray}
P_{\bf q}(\bar{z})
\,=\,\frac{1}{2}\,\left[\,\partial_z\rho\,g\,h_{\bf q
}(\bar{z})\,+\,\partial_z\sigma\,q^2\,h_{\bf q
}(\bar{z})\right]\,\frac{2}{q},
 \label{frozagravcapa}
\end{eqnarray}
where $\partial_z$ stands for $\frac{\partial}{\partial\,z}$.\\
From mean field theories of binary systems
\cite{cahnhilliard,safran} it follows that
$\frac{\partial\,\sigma}{\partial\,z}\,\propto\,\left(\frac{\partial\,\rho}{\partial\,z}\right)^2$.
If the concentration gradient is small, the capillary term becomes
negligible compared to the gravitational one, and we shall drop it
from now on in this case. As before, from
Eq.~(\ref{motointerfacciah2diff}) one may calculate the
correlation function  of bulk fluctuations, recovering the
spectrum calculated and
   commented in Ref. \cite{vailatigiglio98}:\\
\begin{eqnarray}
 \left|h_{{\bf q}\,\omega}(z)\right|^2\,=
 \,{\langle\left| h ({\bf q},z)\right|^2\rangle}_t\,
\frac{ 2\,\frac{ 2\,
\partial_z\rho\,g\,+\,4\,\rho\,\nu\,D\,q^4 }{4\,\rho\,\nu\,q^2}} {
\omega^2\,+\,\left(\frac{
2\,\partial_z\rho\,g\,+\,4\,\rho\,\nu\,D\,q^4
}{4\,\rho\,\nu\,q^2}\right)^2},
 \label{spettrodiffusodiff}
\end{eqnarray}
where the static term is
\begin{eqnarray}
 {\langle\left| h ({\bf q},z_i)\right|^2\rangle }_t\,= \,\frac{1}{A\,{\frac{2}{q}}}\,\frac{k_BT}{\partial_z\rho\,g\,+\,2\rho\,\nu\,D\,q^4}.
 \label{spettrodiffuso2diff}
\end{eqnarray}
  Although these bulk fluctuations have the same origin as capillary
waves, namely velocity fluctuations parallel to the concentration
gradient, the transition from a sharp to a diffuse interface
radically modifies both the static and the dynamic structure
factor of the fluctuations.\\ We shall now outline how one may
evaluate the intensity of light scattered by these fluctuations.
Suppose a plane wave
 having wavevector $K_0$
in vacuum and intensity $I_0$ is propagating in the vertical
direction in a sample having index of refraction $n=n(z)$. The
roughness of the equiconcentration surfaces due to a fluctuation
at $\bar{z}$ introduces a phase dependancy on $(x,y)$. One easily
sees that the resulting scattered intensity per solid angle is
$\frac{dI_{{\bf
q},t}}{d\Omega\,}\,=\,I_0\,\frac{n^2\,K_0^4}{(2\pi)^2}\,\left|\Delta\Phi({\bf
q },t)\right|^2$, where   $\Delta\Phi({\bf q },t)$ is the optical
path variation induced by the fluctuation given by
\begin{eqnarray}
\Delta\Phi({\bf q },t)\,=\,\frac { \partial n } {\partial c
}\,\int dz\,h_z({\bf q },t)\, \,
\partial_z c(z).
 \label{scattint2}
\end{eqnarray}
This integral is similar to that required to calculate the
pressures of Eqs.~(\ref{frozagravcapb}) and (\ref{frozagravcapa})
and,
 depending on the system's local
concentration profile at $\bar{z}$, it can be easily approximated
in the same two limiting cases considered above, sharp interface
or concentration gradient in bulk phase.

  Scattering from the whole sample is given by
integration over the sample thickness of the scattering due to
fluctuations arising in a single layer, weighed with the
probability of being in that layer and with the density of
fluctuations, so that one
 integrates the single fluctuation intensity in
$d\bar{z}\frac{q}{2}$.\\ If the sample comprises both bulk regions
and a sharp interface having a concentration difference $\Delta
c_{int}$, the total scattered light is then:
\begin{eqnarray}
\frac{dI_{{\bf
q},t}}{d\Omega\,}\,=\,I_0\,\frac{n^2\,K_0^4}{(2\pi)^2}\,\left(\frac
{
\partial n } {\partial c }\right)^2\, \left[
 \left( \Delta c_{int}
\right)^2\,\left|h_{z_i}({\bf q },t)\right|^2\,
+\,\int_{bulk}d\bar{z}\frac{2}{q}\left[\partial_z
c(z)\Bigg|_{\bar{z}} \right]^2\left|h_{\bar{z}}({\bf q
},t)\right|^2 \right]\nonumber\\
 \label{scattintb3}
\end{eqnarray}
We have used Eq.~(\ref{scattintb3}) to fit our data after approximating the bulk phase
integral by considering that most of the scattering is due to the layers with the greatest
$\partial_z c(z)$.

%
%
%
%


\begin{references}
\bibitem{Man13}   L.I.Mandelstam,
                {\em Ann. der Physik} {\bf{41}}, 609 (1913).
\bibitem{kating68} R.H.Katyl,and U.Ingard,
                {\em  Phys. Rev. Lett.} {\bf{20}}, 248 (1968).
\bibitem{huwebb69} J.S.Huang, and W.W.Webb,
                {\em Phys. Rev. Lett.} {\bf{23}}, 160 (1969).
\bibitem{benedek71} J.Zollweg, G.Hawkins, and G.B.Benedek,
                {\em Phys. Rev. Lett.} {\bf{27}}, 1182 (1971).
\bibitem{boumeu72} M.A.Bouchiat, and J.Meunier,
                {\em J. de Physique Colloque} {\bf{33}}, C1-141 (1972).
\bibitem{joseph} D.D. Joseph,{\em  Eur. J. Mech., B/Fluids} {\bf 9}, 565 (1990).
\bibitem{maherprl} S.E.May, and J.V.Maher,
                {\em Phys. Rev. Lett.} {\bf{67}}, 2013 (1991).
\bibitem{petitjeans} P. Petitjeans,
                {\em C. R. Acad. Sci. Paris }{\bf 322}, 673 (1996).
\bibitem{maherpre} D.H.Vlad, and J.V.Maher,
                {\em Phys. Rev. E} {\bf{59}}, 476 (1999).
\bibitem{langevin} D.Langevin,
                {\em Light scattering by liquid surfaces and complementary techniques.} (Dekker, New York, 1992).
\bibitem{boumeu71} M.A.Bouchiat, and J.Meunier,
                {\em J. de Physique} {\bf{32}}, 5611 (1971).
\bibitem{cfgpp} M.Carpineti et al.,
                {\em Phys. Rev. A} {\bf{42}}, 7347 (1990).
\bibitem{vailatigiglio96} A.Vailati, and M.Giglio,
                {\em Phys. Rev. Lett.} {\bf{77}}, 1484 (1996).
\bibitem{atackrice}   D.Atack, and O.K.Rice,
                {\em Discuss. Faraday  Soc.} {\bf{15}}, 210 (1953).
\bibitem{vailatigiglio97} A.Vailati, and M.Giglio,
                {\em Nature} {\bf{390}}, 262 (1997).
\bibitem{vailatigiglio98} A.Vailati, and M.Giglio,
                {\em Phys. Rev. E} {\bf{58}}, 4361 (1998).
\bibitem{doriano} D.Brogioli, A.Vailati, and M.Giglio, {\em Phys. Rev.
E} {\bf 61}, R1 (2000).
\bibitem{fischliq} C.Cohen, J.W.H.Sutherland, and J.M.Deutch,
                {\em Phys. Chem. Liq.} {\bf{2}}, 213 (1971).
\bibitem{cahnhilliard}   J.W.Cahn, and J.E.Hilliard,
                {\em J. Chem. Phys.} {\bf{28}}, 258 (1958).
\bibitem{safran} S.A.Safran,
                {\em Statistical thermodynamics of surfaces, interfaces and membranes.} (Addison-Wesley, Reading, 1994).



\end{references}
\end{document}